\documentstyle[preprint,revtex]{aps}
\math-with-secnums
\tightenlines
\begin{document}
\draft
\begin{title}
{\bf SPIN-WAVE THEORY AND}\\
{\bf FINITE-SIZE SCALING FOR}\\
{\bf THE HEISENBERG ANTIFERROMAGNET}
\end{title}
\author{Zheng Weihong$^{*}$ and C.J. Hamer$^{\dagger}$}
\begin{instit}
School of Physics,     \\
The University of New South Wales,        \\
P.O. Box 1, Kensington, NSW 2033, Australia.
\end{instit}
\begin{abstract}
Spin-wave perturbation theory for the Heisenberg antiferromagnet at zero
temperature is used to compute the finite-lattice corrections to the
ground state energy, the staggered magnetization and the energy gap.
The dispersion relation, the spin-wave velocity and the
bulk ground state energy to order $O(1/S^2)$ are also computed for the square
lattice. The results agree very well with the predictions of Neuberger
 and Ziman and Fisher.
\end{abstract}
\pacs{PACS Indices: 75.30.Ds, 75.10.Jm, 05.30.-d, 05.50.+q, 64.60.cn. \\
\\  \\
(Submitted to  Phys. Rev. B, ~~BV4570) }
\newpage

\section{INTRODUCTION}

The Heisenberg antiferromagnet on a square lattice has recently come
under intensive study by a variety of methods, because of its possible
relevance to high-T$_{\rm c}$ superconductors, Reviews of this work have
been given by Barnes \cite{bar} and Manousakis\cite{man}.

The finite-size scaling behaviour of the system has been predicted by
Neuberger and Ziman\cite{neu} and Fisher\cite{fis}, for use in the
analysis of Monte Carlo results. These predictions are based on general
arguments that the large-distance, low-energy behaviour of the system will
 be dominated by massless, soft magnon modes, which can be described by a
simple effective action involving just three unknown parameters, which be
taken as, for instance, the spin-wave velocity $v$, the helicity modulus or
spin-wave stiffness $\rho_s $, and the staggered magnetization $M^+$. The
values of these parameters are not predicted, but must be calculated from
the microscopic Hamiltonian, or fitted to experiment.

Spin-wave perturbation theory has been found to give a comprehensive and
surprisingly accurate description of the Heisenberg antiferromagnet on a
square lattice. Our aim in this paper is to use spin-wave theory to compute
the finite-size scaling behaviour of the system, make comparison with  the
predictions of Neuberger and Ziman\cite{neu} and  Fisher\cite{fis}, and
determine the three parameters referred to above. We also present some
higher-order spin-wave results for the bulk properties of the system.

The spin wave theory for the Heisenberg antiferromagnet was originally
developed by Anderson\cite{and}, and then extended to second
order by Kubo\cite{kub} and
Oguchi\cite{ogu}. The theory was extended to higher order
firstly by Harris et al\cite{har}, and recently by Kopietz\cite{kop},
 Castilla and Chakravarty\cite{cas}, Igarashi and
Watabe\cite{iga}, Canali, Girvin and Wallin\cite{can}, Gochev\cite{goc},
and the present authors\cite{cz}.
The works\cite{har,kop,cas,can,goc} relied on the Dyson-Maleev
transformation, while reference \cite{iga}
used the Holstein-Primakoff  transformation.
Harris et al\cite{har} and Kopietz\cite{kop} studied magnon damping
at low temperature.
Castilla and Chakravarty\cite{cas}  calculated
the staggered magnetization $M^+$ at zero temperature, and concluded that
\begin{equation}
M^+ = S-0.19660  - 0.00068 S^{-2} + O(S^{-3})  ~.
\end{equation}

Igarashi and Watabe\cite{iga} gave results for the renormalization
factor of the
spin-wave velocity $Z_c$, the staggered
magnetization $M^+$, the transverse susceptibility $\chi_{\perp}$,
and the spin-stiffness
constant $\rho_s$ at zero temperature as
\begin{eqnarray}
Z_c &=& 1 + {0.158 \over 2S} - {c_2 \over (2S)^2 },
\quad c_2 \succeq 0.02~, \nonumber \\
M^+ &=& S - 0.197 - {0.01 \over (2S)^2 }~, \nonumber \\
\chi_{\perp} &=& {1\over 8} {\Big [}1 - {0.552 \over 2S} +
{0.04 \over (2S)^2 } {\Big ]}~,  \\
\rho_s &=& \cases {S^2[ 1-{0.236 \over 2S} - {0.15 \over (2S)^2 }], &via
relation $\rho_s = 8 S^2 Z_c^2 \chi_{\perp} $; \cr
     S^2 [1-{0.236 \over 2S} - {0.05 \over (2S)^2 }], &direct calculation,
\cr }
     \nonumber
\end{eqnarray}
where they claimed the inconsistency in $\rho_s$ was due to a rough
estimate of $Z_c$.

Canali, Girvin and Wallin\cite{can} calculated the renormalization
factor of the spin-wave velocity $Z_c$,
\begin{equation}
Z_c = 1 + {0.15795 \over 2S} + {0.0215(2) \over (2S)^2 }
\end{equation}

Gochev\cite{goc} evaluated the ground state energy,
\begin{equation}
E_0/N = -2 S^2 - 0.31588 S - 0.01246 - 0.00210 S^{-1} + O(S^{-2}) ~.
\end{equation}

The present authors\cite{cz} used both the Dyson-Maleev and Holstein-Primakoff
formalisms
to investigate the anisotropic Heisenberg antiferromagnet, and obtained the
consistent
results for the isotropic model:
\begin{eqnarray}
E_0/N &=& -2S^2 - 0.315895 S -0.012474 + 0.000216(6)/S +
 O(S^{-2})~, \nonumber \\
M^+ &=& S - 0.1966019 + 0.000866(25) S^{-2} + O(S^{-3})~,  \\
\chi_{\perp} &=& 0.125 - 0.034447 S^{-1} + 0.001701(3) S^{-2} + O(S^{-3})~.
\nonumber
\end{eqnarray}

Obviously, there are substantial discrepancies between the different
 third order
spin-wave results. We believe
this is probably due to the fact that works\cite{cas,iga,goc}  used the
following relation (in our notation\cite{cz}):
\begin{equation}
\sinh \theta_k = [ (1-\gamma_k^2 )^{-1/2} -1 ]^{1/2}/\sqrt{2}~,
\end{equation}
which holds only if $\gamma_k \geq 0 $. But because there is a
delta function $\delta_{\bf 1+2, 3+4}$ in the vertices ${\cal V}_i^{(0)} $,
the momentum $k$ can not always be inside the first
Brillouin zone, and $\gamma_k $ can be less than 0: in this case,
the above relation
must be replaced by:
\begin{equation}
\sinh \theta_k = - [ (1-\gamma_k^2 )^{-1/2} -1 ]^{1/2}/\sqrt{2} ~.
\end{equation}

In the present work, we calculate the finite-lattice corrections
to the ground-state energy, the staggered magnetization and the mass gap
for the cases of the one-dimensional linear chain and the two-dimensional
square lattice. The results for the square lattice are in good agreement with
the finite-size
scaling predictions and Monte Carlo simulation. For the square
lattice, we also present some further calculations for the third order
spin-wave velocity via both the Dyson-Maleev and Holstein-Primakoff
formalisms, and the fourth order ground state energy via Dyson-Maleev
formalism. For the Holstein-Primakoff formalism, there are some
divergent terms in the third order spin-wave velocity, but they cancel each
other, and the final results are the same as in the Dyson-Maleev
formalism. We will not repeat the derivation of the general spin-wave theory,
but the notations here have the same meanings as in our
previous paper\cite{cz}.

The arrangement of the paper is as follows. In section II we calculate
the finite-lattice corrections. In section III and section IV,
we calculate the third-order spin-wave velocity and the fourth order
ground state energy, respectively. In section V we make comparison with the
prediction of Neuberger and Ziman, and Fisher, and summarize our conclusions.

\section{FINITE-LATTICE CORRECTIONS}
In our previous paper\cite{cz}, we discussed the bulk properties of the model.
Here, we discuss the finite-lattice corrections for the isotropic
 case ($x=1$). The
finite-size scaling corrections can give us a great deal of
information about the model, using either finite-size scaling theory, or
the theory of conformal invariance at criticality in (1+1) dimensions.

\subsection{Ground-state energy and staggered magnetization}

The properties of the isotropic Heisenberg antiferromagnet such as the ground
 state energy $E_0$ and the staggered magnetization $M^+$ are
 functions of $C_n(1)$ which is defined by
\begin{equation}
{C}_n (x) = {2\over N} \sum_{k} {\big [} (1- x^2 \gamma^2_k
)^{n/2} -1 {\big ]}~,
\end{equation}
where the sum over $k$ denotes a sum over the first Brillouin zone of
sublattice $l$.
For a bulk system, the momentum $k$ is continuous over the first
Brillouin zone,
but for a finite-lattice system, the momentum $k$ is discrete.
For the one-dimensional
 linear chain and two-dimensional square-lattice, the structure
 factor $\gamma_k$, the first Brillouin
 zone for a bulk system and the discrete momentum $k$ for a
finite-lattice system are:

1. one-dimensional linear chain:
\begin{eqnarray}
&& \gamma_k = \cos (k_x a)~,  \\
{\rm momentum}~k:&& ~0< k_x a \leq \pi ~~~{\rm bulk ~system} \nonumber \\
&& ~ k_x(i)= {\pi i\over a L_l}, \quad i=1,2,\cdots,L_l \quad
 {\rm finite ~lattice ~system} \nonumber \\
&& L_l=L/2 ~;  \nonumber
\end{eqnarray}

2. two-dimensional square lattice:
\begin{eqnarray}
&& \gamma_k = \cos (\sqrt{2} k_x a/2 ) \cos (\sqrt{2} k_y a/2)~,  \\
{\rm momentum}~k:&& ~0< \sqrt{2} k_x a/2, \sqrt{2} k_y a/2 \leq \pi
 ~~~{\rm bulk ~system} \nonumber \\
&& ~ k_x(i)= {2 \pi i\over \sqrt{2} a L_l},~ k_y(i)=
{2 \pi i\over \sqrt{2} a L_l},\quad i=1,2,\cdots,L_l
\quad {\rm finite ~lattice ~system} \nonumber \\
&& L_l=L/\sqrt{2} ~, \nonumber
\end{eqnarray}
where $L$ and $a$ are the lattice size and the lattice spacing for the
whole system, respectively, and $L_l$ is the lattice size for
sublattice $l$. For convenience, we set the lattice spacing
$a=1$ from now on.

The leading finite-size correction to $C_i$ for the one-dimensional
linear chain can be calculated exactly by using the Euler-Maclaurin
formula\cite{atk}. For a two-dimensional square lattice, the finite-lattice
corrections to $C_i$ can be evaluated by a least-square fit of $C_n (1,L)$
to the form $C_n (1,\infty ) + a/L^{n+2} + b/L^{n+3} + c/L^{n+4} + \cdots $.
The results are:

1. one-dimensional linear chain:
\begin{eqnarray}
C_1 (1)   &=& {2\over \pi }-1 - {2 \pi \over 3L^2}+\cdots ~, \nonumber
\\ && \label{cd1} \\
C_{-1}(1) &=& - {2\over \pi } \ln ({2\pi \over L} )+\cdots ~. \nonumber
\end{eqnarray}

2. two-dimensional square lattice:
\begin{eqnarray}
C_1  (1)  &=& -0.15794742095 - 2.03328/L^3 + 0 \times L^{-4} + O(L^{-5})~,
\nonumber \\
C_{-1}(1) &=& 0.3932039297 - 1.75573607/L + O(L^{-2})~, \label{cd2} \\
C_{-3}(1) &=& 0.20601426 L -0.4488486 + O(L^{-1})~. \nonumber
\end{eqnarray}
Note that the $L^{-4}$ correction to $C_1$ is zero. From equations
(\ref{cd1}) and (\ref{cd2}) we derive the following results.

\subsubsection{One-dimensional lattice}

The ground-state energy, including finite-size correction, is found to be:
\begin{equation}
{E_0\over N} = -S^2 + S({2\over \pi } -1 ) -{1\over 4} ({2\over \pi} -1 )^2
- {\pi \over 3 L^2 } [ 2S - {2\over \pi} +1] + O({1\over S})~.  \label{e01d}
\end{equation}
Note that the energy $m(k)$ of a single-boson state with momentum $k\to 0$ is
\begin{equation}
m(k) = [2 S - {2\over \pi } + 1 + O(1/S)  ] k ~,
\end{equation}
and the spin-wave velocity up to second order is:
\begin{equation}
v=[2 S - {2\over \pi } + 1 + O(1/S)  ]~.  \label{v1d}
\end{equation}
Now according to the theory of conformal invariance, the leading
finite-size correction at the isotropic limit is:
\begin{equation}
{E_0\over N} \sim {E_0(\infty )\over N} - {\pi v c\over 6 L^2 } ~,
\end{equation}
where $c$ is the conformal anomaly, which characterizes the universality
class of the critical point, and the allowed set of critical exponents.
Comparing Eqs. (\ref{e01d}) and  (\ref{v1d}), we
see that the conformal anomaly is obtained as:
\begin{equation}
c=2   ~,
\end{equation}
precisely through second-order in spin-wave expansion. This
 disagrees with the known exact result\cite{woy} $c=1$.

This failure of the spin-wave theory comes as no great
surprise. It has
long been known that spin-wave theory fails qualitatively to
describe the
one-dimensional Heisenberg antiferromagnet: the spin-wave
expansion predicts
a finite staggered magnetization, whereas the exact value\cite{bet}
 is zero; and for integer spin the expansion predicts a zero mass gap,
whereas Haldane's conjecture\cite{hal,aff},
supported by numerical analyses, predicts a finite mass gap.

\subsubsection{Two-dimensional square lattice}

The ground-state energy, including finite-size correction, is found to be:
\begin{equation}
{E_0\over N} = - 2{S^2} - 0.3158948419S -0.01247369389 -{ 4.06656S
\!+\! 0.321151 +O(S^{-1}) \over L^3} +\cdots ~,
\end{equation}
while the staggered magnetization is
\begin{equation}
M^+ =  S -0.196602  + 0.000866(25) S^{-2} + O(S^{-3}) + {{0.8778680+ 0\!
\times \!S^{-1} + O(S^{-2}) }\over L} +\cdots . \label{Mfin}
\end{equation}
This last result must be interpreted with some care. Strictly speaking,
the staggered magnetization at zero magnetic field on a finite lattice is
zero (see later). Equation (\ref{Mfin}) describes the ``bulk value" obtained
 either at a finite but very small field, or else, perhaps, by a measurement
of the mean square magnetization.

A comparison of these results with the predictions of Neuberger and Ziman
will be given in Section V.

\subsection{Energy Gap and Zero Modes}

We now need to take careful consideration of the ``zero modes" of the system,
 which have no special effect
on the calculation of bulk properties, or the finite-lattice corrections to
the ground state energy and bulk staggered magnetization, but which do play
a crucial
role in the finite-lattice behaviour of the energy gap.

Using the Dyson-Maleev representation, one finds after a Fourier
transformation that the terms in the Hamiltonian which involve only
zero-momentum (${\bf k}=0 $) modes  are,
\begin{eqnarray}
H_{{\bf k}=0}=&&zS(a^{\dag}_{\bf 0} a_{\bf 0} +b^{\dag}_{\bf 0} b_{\bf 0} +
a_{\bf 0} b_{\bf 0} + a_{\bf 0}^{\dag} b_0^{\dag} ) \nonumber \\
&& -{z\over N} (a^{\dag}_{\bf 0} a_{\bf 0} a_{\bf 0} b_{\bf 0} +
a^{\dag}_{\bf 0} b^{\dag}_{\bf 0} b^{\dag}_{\bf 0} b_{\bf 0} + 2
a^{\dag}_{\bf 0} a_{\bf 0} b_{\bf 0}^{\dag} b_{\bf 0} ) ~. \label{hk0}
\end{eqnarray}
Now consider the operators corresponding to the total spin on the even ($l$)
and odd ($m$) sublattices:
\begin{eqnarray}
J_1^{\pm}=\sum_l S^{\pm}_l, && \qquad J_1^z=\sum_l S_l^z ~, \nonumber \\ &&
\label{j1j2} \\
J_2^{\pm}=\sum_m S^{\pm}_m, && \qquad J_2^z=\sum_m S_m^z ~.  \nonumber
\end{eqnarray}
If we carry out the same process, representing the spin operators in term of
boson operators, Fourier transforming, and then dropping all terms which
involve non-zero momentum, we find for the rotation-invariant combinations
\begin{eqnarray}
({\bf J}^2_1 )_{{\bf k}=0} && = {NS \over 2} ( {NS \over 2} +1 )~,
 \nonumber \\ && \\
({\bf J}^2_2 )_{{\bf k}=0} && = {NS \over 2} ( {NS \over 2} +1 )~,
 \nonumber \\
({\bf J}_1 + {\bf J}_2 && )^2_{{\bf k}=0} = NS + NS (a^{\dag}_{\bf 0}
a_{\bf 0} +b^{\dag}_{\bf 0} b_{\bf 0} + a_{\bf 0} b_{\bf 0} +
a_{\bf 0}^{\dag} b_0^{\dag} ) \nonumber \\
&& -(a^{\dag}_{\bf 0} a_{\bf 0} a_{\bf 0} b_{\bf 0} + a^{\dag}_{\bf 0}
b^{\dag}_{\bf 0} b^{\dag}_{\bf 0} b_{\bf 0} + 2 a^{\dag}_{\bf 0}
a_{\bf 0} b_{\bf 0}^{\dag} b_{\bf 0} )~. \label{j12}
\end{eqnarray}
Comparing ({\ref{hk0}) with ({\ref{j1j2})-({\ref{j12}), we see that
\begin{eqnarray}
H_{{\bf k}=0} =&&-zS +{z\over N} ({\bf J}_1 + {\bf J}_2 )^2_{{\bf k}=0}
\nonumber \\
&& =-zS + {z\over N} [ 2 ({\bf J}_1^2 + {\bf J}_2^2 ) - ( {\bf J}_1 -
 {\bf J}_2 )^2 ]_{{\bf k}=0}~.
\end{eqnarray}

Thus, if one restricts oneself entirely to the zero-mode sector, the
situation is just that discussed by Neuberger and Ziman\cite{neu}, or
more generally by Fisher and Privman\cite{fipr}: the spins on each
sub-lattice are aligned with each other, so that the total sub-lattice
spins ${\bf J}_1^2$ and ${\bf J}_2^2$ are fixed at their maximum possible
values, while the spins on different sub-lattices are anti-aligned, such
that in the ground state the total spin $({\bf J}_1 +{\bf J}_2)^2 $ is zero.
The Hamiltonian is rotationally symmetric, of course, and the zero-mode
spectrum provides a Wigner representation of this symmetry. In particular,
the ground state eigenvector on the finite lattice is also rotationally
symmetric, so that the order parameter in any particular direction is zero.
The relative energy eigenvalues in the zero-mode sector are
\begin{equation}
\Delta E_{{\bf k}=0} = z j (j+1)/N, \qquad j=0,1,2,\cdots \label{enegap1}
\end{equation}
so that the energy gap to the first excited state is
\begin{equation}
m_N=2z/N~.
\end{equation}

The staggered magnetic field operator is
\begin{eqnarray}
V= &&h ( \sum_l S_l^z - \sum_m S^z_m ) \nonumber \\
=&& h (J_1^z - J_2^z )~.
\end{eqnarray}
If $h=0$, the Hamiltonian is rotationally symmetric, and the spontaneous
magnetization vanishes, as noted above. If $h$ is large enough, then $V$
will dominate over $H_{{\bf k}=0}$, and the ground state will be the
eigenstate with ${\bf J}_1^2 = {\bf J}_2^2 = NS/2 $, $ J_1^z - J_2^z =
-N S $, so that the staggered magnetization will take its bulk value
\begin{equation}
M^+ = - {1\over N} {\partial E_0 \over \partial h} = S~,
\end{equation}
to leading order. The condition on the field strength required for this
to happen is
\begin{equation}
NhS \gg z/N ~,
\end{equation}
i.e.
\begin{equation}
N^2 h \gg z/S ~,
\end{equation}
in agreement with the arguments of Fisher and Privman\cite{fipr}.

A little consideration shows that the neglected terms involving non-zero
momentum modes will modify these results, producing corrections of higher
order in spin-wave perturbation theory, which we have not explicitly
 calculated. The basic scenario will remain the same, however. We have
also checked that the same results hold in  the Holstein-Primakoff
 representation, to leading order.

\section{DISPERSION RELATION AND SPIN-WAVE VELOCITY}
In our previous paper\cite{cz}, we calculated the spin-wave
energy of a
single boson state with momentum $k=0$ for the anisotropic Heisenberg
antiferromagnet. Here we calculate the spin wave energy $m(k)$ of a single
boson state with non-zero momentum $k$, and then estimate the spin-wave
velocity $Z_c$ and the stiffness constant $\rho_s$. For the anisotropic
Heisenberg antiferromagnet ($x\not= 1$), the spin
wave velocity is zero, so in this section we only consider the isotropic
model ($x=1$),  using both the Dyson-Maleev and
Holstein-Primakoff formalisms.

\subsection{Dyson-Maleev Formalism}
As before\cite{cz}, the spin-wave energy of a single-boson state with
momentum $k$ up to order
$O(1/S)$ can be derived as:
\begin{equation}
m(k)=m^{(1)}(k) + m^{(0)}(k) + m^{(-1)}(k) + O(1/S^2) ~,
\end{equation}
where
\begin{eqnarray}
m^{(1)}(k) &&= zS(1-x^2 \gamma^2_k )^{1/2}~, \nonumber \\
m^{(0)}(k) &&=-{z\over 2} {\Big [} (1-x^2 \gamma^2_k )^{1/2} C_1
+(1-x^2) (C_{-1} -C_1 )\gamma^2_k (1-x^2 \gamma_k^2 )^{-1/2} {\Big ]}~, \\
m^{(-1)}(k) &&= \Delta m_a^{(-1)}(k) + \Delta m_b^{(-1)}(k)
+ \Delta m_c^{(-1)}(k) + \Delta m_d^{(-1)}(k) + \Delta m_e^{(-1)} (k)~,
\nonumber
\end{eqnarray}
and $\Delta m_a^{(-1)}(k) $ is the contribution from Fig.2a in the
 previous paper \cite{cz}, etc:
\begin{eqnarray}
\Delta m_a^{(-1)}(k)&=&-{z\over 8x^2 S} (1-x^2)^2 (C_{-1} - C_1 )^2
\gamma_k^2 (1-x^2 \gamma^2_k )^{-3/2}~, \nonumber \\
\Delta m_b^{(-1)}(k)&=&-{z\over 2S} {\Big (}{2\over N}{\Big )}^2
\sum_{i=1}^3
\delta_{{\bf 1}+{\bf 2},{\bf 3}+k } { V_5^{(0)}({\bf 1},{\bf 2},{\bf 3},k)
V_5^{(0)}({\bf 3},k,{\bf 1},{\bf 2}) \over \sum_{i=1}^3 (1-x^2
\gamma^2_i )^{1/2}
+ (1-x^2 \gamma^2_k )^{1/2} }~, \nonumber \\
\Delta m_c^{(-1)}(k)&=& - {z \over 2 S} {\Big (}{2\over N}{\Big )}^2
 \sum_{i=1}^3
\delta_{{\bf 1}+{\bf 2},{\bf 3}+k } {
V_2^{(0)}({\bf 1},{\bf 2},{\bf 3},k)
V_3^{(0)}({\bf 3},k,{\bf 1},{\bf 2})
\over \sum_{i=1}^3 (1-x^2 \gamma^2_i )^{1/2}
- (1-x^2 \gamma^2_k )^{1/2} }~,  \\
\Delta m_d^{(-1)}(k)&=&-{z\over 8xS} ({2 \over N}) \sum_{k'} (1\!-\!x^2)
(C_{-1}\! -\!C_1) \gamma_{k'} (1\!-\!x^2 \gamma^2_{k'} )^{-1}
[ V_2^{(0)} (k',k,k,k') \!+\! V_2^{(0)}(k,k',k,k') ]~, \nonumber \\
\Delta m_e^{(-1)}(k)&=&-{z\over 8xS} ({2 \over N}) \sum_{k'} (1\!-\!x^2)
(C_{-1} \!-\!C_1) \gamma_{k'} (1\!-\!x^2 \gamma^2_{k'} )^{-1}
[ V_3^{(0)} (k',k,k,k') \!+\! V_3^{(0)} (k',k,k',k) ]~, \nonumber
\end{eqnarray}
with
\begin{equation}
\Delta m_d^{(-1)}(k) + \Delta m_e^{(-1)}(k)={z\over 4 x^2 S} (1-x^2)^2
(C_{-1}-C_{1} )\gamma^2_k (1-x^2 \gamma^2_k )^{-1/2} ~.
\end{equation}
In the isotropic limit ($x=1$),
\begin{equation}
\Delta m_a^{(-1)} (k) = \Delta m_d^{(-1)}(k) + \Delta m_e^{(-1)}(k)= 0~,
\end{equation}
and
\begin{equation}
\Delta m_b^{(-1)}(k) + \Delta m_c^{(-1)}(k)= -{z\over 2S}
(1- \gamma_k^2 )^{1/2} m_{bc}(k) ~,
\end{equation}
where
\begin{eqnarray}
m_{bc}(k) =&& {\Big (}{2\over N}{\Big )}^2 \sum_{k_1,k_2,k_3}
{ \delta_{{\bf 1}+{\bf 2}, {\bf 3}+k}
\over [\sum_{i=1}^3 (1-\gamma^2_i )^{1/2}]^2 - [(1-\gamma^2_k)^{1/2}]^2 }
\times \nonumber \\
&& {\Big \{} { [
V_5^{(0)} ({\bf 1},{\bf 2},{\bf 3},k) V_5^{(0)}({\bf
3},k,{\bf 1},{\bf 2}) \!+ \!
V_2^{(0)}({\bf 1},{\bf 2},{\bf 3},k) V_3^{(0)}({\bf
3},k,{\bf 1},{\bf 2})
] \sum_{i=1}^3 (1\!-\! \gamma^2_i )^{1/2} \over
(1- \gamma^2_k )^{1/2} }  \nonumber \\
&& + [ V_2^{(0)}({\bf 1},{\bf 2},{\bf 3},k)
V_3^{(0)}({\bf 3},k,{\bf 1},{\bf 2}) - V_5^{(0)} ({\bf
1},{\bf 2},{\bf 3},k)
V_5^{(0)}({\bf 3},k,{\bf 1},{\bf 2}) ]  {\Big \} } ~.
\end{eqnarray}

For the square lattice, the integration of ${m}_{bc}(k) $ can be evaluated
for finite lattices up to
lattice size $L_l=120$, then extrapolated to the infinite lattice by the
form ${m}_{bc}(k,\infty)+ {a\over
L^2 } + {b\over L^3} + {c\over L^4} $. Fig. 1 shows the dispersion
relation along the
line $k_x=k_y$. The third order results here disagree with
those of Igarashi
and Watabe\cite{iga}.

In the limit $k\to 0$, using MATHEMATICA, ${m}_{bc}(k) $ can be found to
have two integration parts:
\begin{equation}
{m}_{bc}(k)= {\Big (}{2\over N}{\Big )}^2 \sum_{i=1}^3 \delta_{{\bf 1}+{\bf 2},
{\bf 3}+k} [ 2 {\tilde{m}}_{bc}^{(-1)}/k+ {\tilde{m}}_{bc}^{(0)} ]~,
\quad k\to 0 ~,
\end{equation}
where the expressions for ${\tilde{m}}_{bc}^{(-1)}$ and
${\tilde{m}}_{bc}^{(0)}$  are too complicated to be given here. The
integral of the divergent part is found to be zero. Replacing all
 $(1-\gamma_i^2)^{1/2}$ $(i=1,2,3)$ by $(1-t^2 \gamma_i^2)^{1/2} $ in the
 second integration
part, and using the series expansion technique in our previous
paper\cite{cz},
we can obtain a series in $t$ for the integral of ${m}_{bc} (0)$
for the infinite lattice. This series can be supplied on request.
Extrapolating \cite{cz} the series to the limit $t\to 1$, we
get:
\begin{equation}
{m}_{bc}(0) = -0.01076(1) ~.
\end{equation}

Therefore, the energy gap of the isotropic Heisenberg antiferromagnet
at the small $k$ limit is:
\begin{equation}
m(k)=4S {\Big [} 1+ {0.157947421 \over 2S} + {0.02152(2) \over (2S)^2 }
+ O(1/S^3) {\Big ]} (\sqrt{2} k/2)   ~,
\end{equation}
and the renormalization factor of the spin-wave velocity is:
\begin{equation}
Z_c={v\over v_0}= 1 + {0.157947421 \over 2 S} + { 0.02152(2) \over (2S)^2 }
+O(1/S^3) ~,
\end{equation}
where $v_0$ is the ``bare" spin-wave velocity obtained in the linear
spin-wave approximation, namely, $v_0=2\sqrt{2} S$.
The stiffness constant $\rho_s$ can be estimated by using
the hydrodynamic relation:
\begin{equation}
\rho_s = v^2 \chi_{\perp} = S^2 {\Big [} 1-{0.23525 \over 2S} -{0.0517(2)
\over (2S)^2 } + O(1/S^3) {\Big ]} ~,
\end{equation}
where $\chi_{\perp}$ is the uniform perpendicular susceptibility\cite{cz}.

Here the result for the spin-wave velocity is different from that of Igarashi
and Watabe\cite{iga}, but agrees with that of Canali, Girvin,  and
 Wallin \cite{can}, and
the stiffness constant $\rho_s$ is consistent with the direct
calculation of Igarashi and Watabe.

\subsection{Holstein-Primakoff Formalism}
The energy gap $m(k)$ can be calculated using the same
method as in the Dyson-Maleev formalism, and the result is:
\begin{equation}
m(k)= m^{(1)}(k)+ m^{(0)}(k) + m^{(-1)}(k) + O(1/S^2) ,
\end{equation}
where  $m^{(1)}(k)$ and $ m^{(0)}(k) $ are the same as in the Dyson-Maleev
formalism, and $ m^{(-1)}(k) $ is:
\begin{equation}
m^{(-1)}(k)=\Delta m_0^{(-1)}(k)+\Delta m_a^{(-1)}(k)+\Delta m_b^{(-1)}(k)+
\Delta m_c^{(-1)}(k)+\Delta m_d^{(-1)}(k)+\Delta m_e^{(-1)}(k)~,
\end{equation}
where
\begin{equation}
\Delta m_0^{(-1)}(k)={z\over 32S}{\Big \{}2(C_{-1}\!-\!C_1)(C_{-1}\!+\!1)
\!+\!x^2 \gamma_k^2 [C_{-1} (C_{-1}\!+\!2)\!-\! {3\over x^2}
 (C_{-1}\!-\!C_1)^2]{\Big \} } (1\!-\!x^2\gamma_k^2 )^{-1/2}~,
\end{equation}
while the results for $\Delta m_a^{(-1)}(k)$ and  $ \Delta m_d^{(-1)}(k)+
\Delta m_e^{(-1)}(k) $ are the same as in the Dyson-Maleev
formalism for all bipartite lattices. The terms $\Delta m_b^{(-1)}(k)$,
$ \Delta m_c^{(-1)}(k) $ and ${m}_{bc}(k)$ have the same expression as in
the Dyson-Maleev
formalism except  that the vertices  $ V_i^{(0)}$ are the
Holstein-Primakoff vertices.

In the isotropic limit $x=1$:
\begin{eqnarray}
\Delta m_0^{(-1)}(k)=&&{z\over 32S}{\Big \{}[2(C_{-1}-C_1)(C_{-1}+1)
+ C_{-1} (C_{-1}+2) - 3(C_{-1}-C_1)^2 ] (1-\gamma_k^2)^{-1/2} \nonumber \\
&&+[3(C_{-1}-C_1)^2-C_{-1} (C_{-1}+2 )] (1-\gamma^2_k )^{1/2} {\Big \} }~.
\end{eqnarray}

For the square lattice, the integration  $ {m}_{bc}(k) $ can
also be evaluated for a finite-lattice up to
lattice $L_l=120$, while here the finite lattice correction is ${a\over
L } + {b\over L^2} + \cdots $. The calculation gives the
same dispersion relation as Fig. 1.

In the limit $k\to 0$,
\begin{eqnarray}
\Delta m_0^{(-1)}(k)=&&{z\over 32S}{\Big \{}[2(C_{-1}-C_1)(C_{-1}+1)
+ C_{-1} (C_{-1}+2) - 3(C_{-1}-C_1)^2 ] {2\over \sqrt{2} k}\nonumber \\
&& +[3(C_{-1}-C_1)^2-C_{-1} (C_{-1}+2 )] \sqrt{2} k/2 {\Big \} }  \\
=&& {2 \over \sqrt{2} S} [ 0.19568/k - 0.001857 k ], \quad k\to 0 ~.
 \nonumber
\end{eqnarray}
Note that $\Delta m_0^{(-1)}(k)$ is divergent in the limit $k\to 0$,
and ${m}_{bc} (k)$ is found, via MATHEMATICA, to have three
integration parts in the small $k$ limit:
\begin{equation}
{m}_{bc} (k)= {\Big (}{2\over N}{\Big )}^2 \sum_{i=1}^3
 \delta_{{\bf 1}+{\bf 2},
{\bf 3}+k} {\big [}  4 {\tilde{m}}^{(-2)}_{bc} /k^2 +
2 {\tilde{m}}^{(-1)}_{bc} /k + {\tilde{m}}^{(0)}_{bc}  {\big ]},
\quad k\to 0 ~.
\end{equation}
Let
\begin{equation}
{m}^{(i)}_{bc} ={\Big (}{2\over N}{\Big )}^2 \sum_{i=1}^3
\delta_{{\bf 1}+{\bf 2},
{\bf 3}+k} {\tilde{m}}^{(i)}_{bc},\quad (i=-2,-1,0) ~,
\end{equation}
the integration  ${m}^{(-1)}_{bc} $ is found to be zero, and the integration
 for
${m}^{(-2)}_{bc} $ and ${m}^{(0)}_{bc} $ can be carried out in the same way
as before:
\begin{eqnarray}
{m}_{bc}^{(-2)}= 0.049(2)~, \nonumber \\
{m}_{bc}^{(0)}= -0.012(2)~.
\end{eqnarray}
Therefore, in the limit $k\to 0$, the divergent parts of $\Delta
m_0^{(-1)}(0)$ and $\Delta m_b^{(-1)}(0)+ \Delta m_c^{(-1)}(0) $
cancel each other, and the final result for $m(0)$ is finite,
agreeing with that obtained via the Dyson-Maleev formalism.

\section{FOURTH-ORDER SPIN WAVE RESULTS}
In this section, we present the fourth order spin wave expansion for
the ground state energy within
the Dyson-Maleev formalism. We only consider the case without
an external magnetic field, that is, $h_1=h_2=0$.

According to the Hamiltonian $H$ in Eq.(2.7) of our previous
paper\cite{cz}, there are seven perturbation diagrams shown
in Fig. 2 contributing to the order $O(S^{-2})$ for ground state
energy $E_0$; the first two diagrams also contribute to the
order $O(S^{-1})$ of $E_0$, and the $O(S^{-1})$ part has been
considered in our previous paper. The contributions from each diagram are:
\begin{eqnarray}
\Delta { E}_a&& =  \sum_k { [V_0^{(0)}]^2 \over
2 \{ -zS q_k\! -\! {z\over 2} {\big [}q_k C_1 \!+\!(1\!-\!x^2)
\gamma^2_k q_k^{-1} (C_{-1} \! - \! C_1 ){\big ]} \} } \nonumber \\
&& \equiv \Delta { E}_a^{(-1)} + \Delta { E}_a^{(-2)} + O(S^{-3}) ~,
         \nonumber \\
\Delta { E}_b &&=  - {z^2N \over 8} \left( {2\over N}
 \right) ^3 \sum_{k_i} \delta_{{\bf 1}+{\bf 2},{\bf 3}+{\bf 4}}
{  V_5^{(0)}
( {\bf 1,2,3,4} )  V_5^{(0)}({\bf 3,4,1,2}) \over
\sum_i \{  zS q_i + \! {z\over 2} {\big [} q_i C_1 \!+
\!(1\!-\!x^2) \gamma^2_i q_i^{-1} (C_{-1}\!-\!C_1 ){\big ]} \} }
\nonumber \\
&& \equiv \Delta { E}_b^{(-1)} + \Delta { E}_b^{(-2)} + O(S^{-3}) ~,
         \nonumber \\
\Delta { E}_c^{(-2)}&& =  -{N\over 16zS^2} \left( {2\over N}
\right) ^2 \sum_{k_1, k_2}
[ V_4^{(0)} ({\bf 1},{\bf 2},{\bf 2},{\bf 1}) + V_4^{(0)} ({\bf 2},{\bf 1},
{\bf 1},{\bf 2}) ] V_0^{(0)} ({\bf 1}) V_0^{(0)} ({\bf 2})/(q_{{\bf 1}}
q_{{\bf 2}}) ~, \label{eqE} \\
\Delta { E}_d^{(-2)}&& =   -{N\over 16zS^2} \left( {2\over N}
 \right) ^2 \sum_{k_1, k_2}
[ V_5^{(0)} ({\bf 1},{\bf 2},{\bf 1},{\bf 2}) + V_5^{(0)} ({\bf 1},{\bf 2},
{\bf 2},{\bf 1}) ] V_0^{(0)} ({\bf 1}) V_0^{(0)} ({\bf 2})/(q_{{\bf 1}}
q_{{\bf 2}})~,  \nonumber \\
\Delta { E}_e^{(-2)} && =  - {N\over 4S^2} \left( {2\over N} \right) ^3 \!
\sum_{k_i} \delta_{{\bf 1}+{\bf 2},{\bf 3}+{\bf 4}} {  V_0^{(0)} ({\bf 1})
[ V_3^{(0)} ({\bf 2},\!{\bf 1},\!{\bf 3},\!{\bf 4}) V_5^{(0)} ({\bf 3},
\!{\bf 4},\!{\bf 1},\!{\bf 2}) \! + \!V_2^{(0)} ({\bf 3},\!{\bf 4},\!{\bf 1},
\!{\bf 2}) V_5^{(0)} ({\bf 1},\!{\bf 2},\!{\bf 3},\!{\bf 4}) ]
\over q_{{\bf 1}} (q_{{\bf 1}} + q_{{\bf 2}}+q_{{\bf 3}}+q_{{\bf 4}} ) }~,
     \nonumber \\
\Delta {E}_f^{(-2)} &&= -{zN\over 8S^2}\left( {2\over N}
\right) ^4 \sum_{k_i} \delta_{{\bf 1}+{\bf 2},{\bf 3}+
{\bf 4}}\delta_{{\bf 1}+{\bf 2},{\bf 5}+{\bf 6}}
{V_5^{(0)} ({\bf 1},{\bf 2},{\bf 3},{\bf 4}) V_1^{(0)}
({\bf 3},{\bf 4},{\bf 5},{\bf 6})
V_5^{(0)} ({\bf 5},{\bf 6},{\bf 1},{\bf 2}) \over
(q_{{\bf 1}} + q_{{\bf 2}}
+q_{{\bf 3}}+q_{{\bf 4}} ) (q_{{\bf 1}} + q_{{\bf 2}}+
q_{{\bf 5}}+q_{{\bf 6}}
) } ~, \nonumber \\
\Delta {E}_g^{(-2)} &&= -{zN\over 4S^2}\left( {2\over N}
\right) ^4 \sum_{k_i} \delta_{{\bf 1}+{\bf 2},{\bf 3}+{\bf
4}}\delta_{{\bf 4}+{\bf 6},{\bf 5}+{\bf 1}}
{V_5^{(0)} ({\bf 1},{\bf 2},{\bf 3},{\bf 4}) V_4^{(0)}
({\bf 4},{\bf 6},{\bf 5},{\bf 1})
V_5^{(0)} ({\bf 3},{\bf 5},{\bf 2},{\bf 6})  \over
(q_{{\bf 1}} + q_{{\bf 2}}
+q_{{\bf 3}}+q_{{\bf 4}} ) (q_{{\bf 2}} + q_{{\bf 3}}+q_{{\bf
 5}}+q_{{\bf 6}}
) }~, \nonumber
\end{eqnarray}
where
\begin{eqnarray}
q_k &=& (1-x^2 \gamma^2_k )^{1/2}~, \nonumber \\
\Delta { E}_a^{(-1)} &=& -{zN \over 16 x^4 S} (1\!-\!x^2)^2
( C_{-1}\!-\!C_1 )^2 (C_{-3} \!- \!C_{-1} ) \nonumber \\
\Delta { E}_a^{(-2)} &=&  {zN \over 32 x^6 S^2} (1\!-\!x^2)^2
( C_{-1}\!-\!C_1 )^2
\{ C_1 (C_{-1}\!-\!C_{-3})  \nonumber \\
&&~~~~ +  (1\!-\!x^2) [ C_{-1} (C_{-3}\!-\!C_{-1} ) \!+\!
(C_{-3}\!-\!C_{-5}
) (C_{-1}\!-\! C_1 )] \}  ~, \\
\Delta { E}_b^{(-1)} &=&  - {zN \over 8S} \left( {2\over N}
 \right) ^3 \sum_{k_i} \delta_{{\bf 1}+{\bf 2},{\bf 3}+{\bf 4}}
{  V_5^{(0)}
( {\bf 1,2,3,4} )  V_5^{(0)}({\bf 3,4,1,2}) \over q_{{\bf 1}} +
q_{{\bf 2}}+q_{{\bf 3}}+q_{{\bf 4}}  } ~, \nonumber \\
\Delta { E}_b^{(-2)} &=&  - {zN \over 16S^2} \left( {2\over N}
 \right) ^3 \! \sum_{k_i} \delta_{{\bf 1}+{\bf 2},{\bf 3}+{\bf 4}}
{  V_5^{(0)}
( {\bf 1,\!2,\!3,\!4} ) V_5^{(0)}({\bf 3,\!4,\!1,\!2}) \{ \sum_i
[ q_i C_1 \! +\! (1\!-\!x^2) \gamma^2_k q_i^{-1} (C_{-1} \!-\! C_1)] \}
\over ( q_{{\bf 1}} + q_{{\bf 2}}+q_{{\bf 3}}+q_{{\bf 4}} )^2}~. \nonumber
\end{eqnarray}

At the isotropic limit $x=1$, we can easily prove that:
\begin{eqnarray}
\Delta { E}_b^{(-2)} &&= {C_1 \over 2S} \Delta { E}_b^{(-1)}~,
\nonumber \\ && \label{eer} \\
\Delta { E}_a^{(-1)} &&= \Delta { E}_a^{(-2)}= \Delta { E}_c^{(-2)}
=\Delta { E}_d^{(-2)}=\Delta { E}_e^{(-2)}=0 ~. \nonumber
\end{eqnarray}

Therefore, the ground state energy per site $E_0/N$ is:
\begin{eqnarray}
E_0/N= && -{zS\over 2} \left[ S - C_1 + {1\over 4S} \left[ C_1^2 +
{1-x^2 \over x^2 } (C_{-1}-C_1 )^2 \right] \right] \nonumber \\
&& - {z \over 16 x^4 S} (1-x^2 )^2 (C_{-1} -C_1 )^2 (C_{-3} -C_{-1} )
+ \Delta E_b^{(-1)}/N  \\
&& + (\Delta E_a^{(-2)} \!+ \!\Delta E_b^{(-2)}\! + \!\Delta E_c^{(-2)} \!
+ \!\Delta E_d^{(-2)} \!+\! \Delta E_e^{(-2)} \!+ \!\Delta E_f^{(-2)} \!+ \!
\Delta E_g^{(-2)} )/N \!+ \! O(S^{-3})~.  \nonumber
\end{eqnarray}

Hitherto, the results have been applicable to all bipartite lattices.
We now restrict ourselves to
the two-dimensional square lattice, where $\Delta { E}_c^{(-2)}$ and
$\Delta { E}_d^{(-2)}$ are four dimensional integrals
over the first Brillouin zone: the integrations can be carried out
analytically, and the results are:
\begin{equation}
\Delta { E}_c^{(-2)} + \Delta { E}_d^{(-2)} = - {N \over 8 x^6 S^2}
(1-x^2)^3 (C_{-1}-C_1)^2 (C_{-3}-C_{-1})^2  ~.
\end{equation}

$\Delta { E}_b^{(-2)}$ and $\Delta { E}_e^{(-2)}$ are both six
dimensional integrals
over the first Brillouin zone, while $\Delta { E}_f^{(-2)}$ and
$\Delta { E}_g^{(-2)}$ are eight
dimensional integrals. They have been calculated using two
different methods. The first one is a series expansion in $x$,
  the results can be supplied on request.
Thus, for the spin-${1\over 2}$ model, the series for $E_0/N$ in $x$
from the fourth order spin-wave theory is:
\begin{equation}
{E_0^{{\rm 4th}} \over N}= -{1\over 2} -{85 x^2 \over 512}
- 0.0029749550 x^4 +
   0.00065351749 x^6 + 0.0002705726 x^8 +
   0.000010691 x^{10} + O(x^{12}) ~,
\end{equation}
clearly, this series is closer to the exact series than that for
third-order spin-wave theory\cite{cz}.

The most interesting thing here is the ground state energy at the
isotropic limit ($x=1$).
The series obtained seems to be too short to give a reliable extrapolation
to $x=1$, but
we can also use another technique: the finite-lattice technique
discussed in the
first section. We evaluate numerically Eq.(\ref{eqE}) by dividing
the first Brillouin zone
into a finite number of meshes, $L_l\times L_l $, and extrapolate the
sum to $L_l \to \infty $. For small $x$, this technique confirms the
 results of the
 above series expansion. For $x$ close to 1, the results for
$\Delta { E}_f^{(-2)} $
are shown in Fig. 3 as an example. Extrapolating to $x=1$ and $L_l
\to \infty$, we can get:
\begin{eqnarray}
\Delta { E}_b^{(-1)}/N &&={0.0004285(4) \over 2S} ~, \nonumber  \\
\Delta { E}_f^{(-2)}/N &&= {0.0001054(4) \over S^2} ~,  \\
\Delta { E}_g^{(-2)}/N &&= {3.61(4)\times 10^{-5} \over S^2}~,  \nonumber
\end{eqnarray}
where the results for $\Delta { E}_b^{(-1)}/N$ are consistent
with our previous results\cite{cz}, but substantially more
accurate. The results for $\Delta { E}_b^{(-2)}$ can be found by
using the relation (\ref{eer}):
\begin{equation}
\Delta { E}_b^{(-2)}/N = - { 6.768(6) \times 10^{-5} \over (2S)^2 } ~.
\end{equation}

Therefore, we conclude that for the square lattice
\begin{equation}
E_0/N= -2S^2 - 0.315894842 S - 0.0124736939 + 0.0002142(2)/S + 1.246(9)
\times 10^{-4}/S^2 + O(S^{-3}).
\end{equation}
If we use Pad\'{e} approximants\cite{gut} to analyze the
above series, we get
\begin{equation}
E_0/N=\cases {-0.6693(2) , &$S={1\over 2}$; \cr
              -2.32801(4), &$S=1$. \cr }
\end{equation}

\section{SUMMARY AND CONCLUSIONS}

As further results of spin-wave perturbation theory for
the Heisenberg antiferromagnet, we have calculated the finite-lattice
corrections to the ground state energy and the staggered magnetization, and
also the finite-lattice energy gap,
for both the one-dimensional linear chain
 and two-dimensional
square lattice. We have also calculated the spin-wave velocity, and the ground
state energy to order $O(1/S^2)$ for the square lattice.

For the one-dimensional linear chain, spin-wave theory
gives the conformal anomaly as
\begin{equation}
c=2
\end{equation}
precisely, through second order in the expansion,
whereas the true value\cite{woy}
is $c=1$. This is no great surprise, since it is already well
known that spin-wave theory fails to describe the one-dimensional
chain, incorrectly predicting a non-zero staggered
magnetization, and a mass gap which vanishes for all spins.

For the square lattice of size $N=L^2$, our results may be summarized as
 follows:

Bulk ground state energy per site:
\begin{eqnarray}
e_{\infty}=&& \lim_{N\to \infty} {E_0\over N} = -2S^2  - 0.315894842 S -
 0.0124736939 \nonumber \\
&& + {0.0002142(2)\over S} + {0.0001246(9) \over S^2} + O(S^{-3})~,
\end{eqnarray}
with finite-lattice correction:
\begin{equation}
{E_0\over N}-e_{\infty} = -{ 4.06656S \!+\! 0.321151 + O(S^{-1})
 \over L^3} +\cdots ~.\label{efin}
\end{equation}

Bulk staggered magnetization:
\begin{equation}
M^+_{\infty}=S-0.1966019 + 0.000866(25)S^{-2} + O(S^{-3})~,
\end{equation}
with ``finite-lattice correction"
\begin{equation}
M^+_{N}-M^+_{\infty}={{0.8778680+ 0\! \times \!S^{-1} + O(S^{-2}) }\over L}
 +\cdots ~.
\end{equation}

Transverse susceptibility\cite{cz}:
\begin{equation}
\chi_{\perp} = {1\over 8} - 0.034447 S^{-1} + 0.001701(3) S^{-2}
 + O(S^{-3})~. \label{chisw}
\end{equation}

Dispersion relation in the small $k$ limit:
\begin{equation}
E(k)=2\sqrt{2} S [1 + {0.157947421 \over 2 S} + { 0.02152(2)
\over (2S)^2 } +O(S^{-3}) ] k ~.
\end{equation}

Finite-lattice energy gap:
\begin{equation}
m_N = [8+ O(S^{-1})]/L^2 ~. \label{mfin}
\end{equation}

Spin-wave velocity renormalization factor:
\begin{equation}
Z_c =E(k)/(2\sqrt{2} S k)=  1 + {0.157947421 \over 2 S} + { 0.02152(2)
\over (2S)^2 } +O(S^{-3})~.
\end{equation}

Spin-stiffness constant $\rho_s$:
\begin{equation}
\rho_s = v^2 \chi_{\perp} =S^2 {\Big [} 1-{0.23525 \over 2S} -{0.0517(2)
 \over (2S)^2 } + O(S^{-3}) {\Big ]} ~.
\end{equation}

The spin-wave velocity $Z_c$ was calculated through both the Dyson-Maleev
and Holstein-Primakoff transformations, and the fourth order ground state
 energy via Dyson-Maleev formalism. As before \cite{cz}, in the
 Holstein-Primakoff formalism there are some divergent terms, but the
 divergences eventually cancel one another. Our result for the spin-wave
 velocity is different from that of Igarashi and Watabe\cite{iga}, but
 agrees with that of Canali, Girvin,
and Wallin \cite{can}, and the spin-stiffness constant is consistent with
 the direct calculation of Igarashi and Watabe\cite{iga}.  Obviously, the
corrections of high orders are pretty small, and the spin-wave
theory continues to give consistent results, and improved
convergence towards the exact values.
These results should be compared with other estimates: for the
 spin-${1\over 2}$
model, Runge\cite{run} finds
that $E_0/N=-0.66934(3)$ and $Z_c=1.10(3)$ using a Green's
function Monte
Carlo method, while Singh\cite{sin} predicts that $Z_c=1.18(2)$ and
 $\rho_s=0.73(4)$ using a series expansion.

The finite-lattice predictions from the effective theory of Neuberger
 and Ziman\cite{neu}, and
Fisher\cite{fis} amount to six relations, which are
\begin{eqnarray}
&& {E_0\over N}-e_{\infty} = {2\beta v \over L^3} ~, \label{nze} \\
&& M^+_{\infty}= 2 \kappa_1 \kappa_2 ~, \\
&& M^+_{N}-M^+_{\infty}= - {\kappa_1 \alpha \over 2 \kappa_2 L} ~, \\
&& \chi_{\perp} = 2 \kappa_2^2/v ~, \\
&& E(k)=vk ~, \\
&& m_N={1\over \chi_{\perp} L^2 } ~, \label{nzm}
\end{eqnarray}
involving two calculable structure constants $\alpha=-0.6208 $,
 $\beta=-0.7186$, and three unknown microscopic parameters $\kappa_1$,
 $\kappa_2$ and $v$. Comparing (\ref{nze})-(\ref{nzm}) with
 (\ref{efin})-(\ref{mfin}), we find these predictions are exactly
 satisfied, order by order in spin-wave perturbation theory as far as
 have calculated, with $\alpha=-0.6207464 $ , $\beta=-0.71887$, the
 velocity $v$ given by
\begin{equation}
v={E(k) \over k}= 2\sqrt{2} S [1 + {0.157947421 \over 2 S} + { 0.02152(2)
\over (2S)^2 } +O(S^{-3}) ] ~,
\end{equation}
and
\begin{eqnarray}
\kappa_1 &=& S^{1/2} { \Big [} 2^{1/4} - {0.116900\over S} -
{0.00410(2)\over S^2} + O(S^{-3})  {\Big ]}~, \nonumber \\  && \\
\kappa_2 &=& S^{1/2} {\Big [} 2^{-5/4} - {0.0413305\over S} -
{0.002615(8) \over S^2} + O(S^{-3}) {\Big ]}~. \nonumber
\end{eqnarray}
This agreement is very satisfying, and helps to give us confidence that
 the spin-wave results are correct.

The finite-lattice energy gap has been measured by Carlson \cite{car} using
 a Green's function Monte Carlo method, with result
\begin{eqnarray}
&& \Delta E(j=1) \simeq 10/L^2 \qquad {\rm for ~total ~spin\!-\!1}
 \nonumber \\ && \\
&& \Delta E(j=2) \simeq 29/L^2 \qquad {\rm for ~total ~spin\!-\!2} \nonumber
\end{eqnarray}
while from the data of Runge \cite{run} we obtain
\begin{eqnarray}
&& \Delta E \simeq 5.2j(j+1)/L^2 \qquad {\rm for~~} L=6\nonumber \\ && \\
&& \Delta E \simeq 5.4j(j+1)/L^2 \qquad {\rm for~~} L=8\nonumber
\end{eqnarray}
Similar results can be found in the paper by Gross et al \cite{gro}. The
 leading
order spin-wave results (\ref{enegap1}) is
\begin{equation}
\Delta E = { 4 j (j+1) \over L^2}~.
\end{equation}
If we use the relation (\ref{nzm}) together with (\ref{chisw}), the higher
 order result is predicted to be
\begin{eqnarray}
\Delta E = && [4 + 1.102304/S + 0.2493(1)/S^2 + O(S^{-3})] {j (j+1)
 \over L^2}, \nonumber \\
\simeq  && 7.202  j (j+1)/L^2 ~,
\end{eqnarray}
although this has not been confirmed by direct calculation. The result is
 at least in the same ballpark as ``experiment".

\acknowledgments

This work forms part of a research project supported by a grant
from the Australian Research Council.

\newpage
\figure{The spin-wave energy $m(k)$ as a function of  momentum
$ak_x/\sqrt{2}$ along a line $k_x=k_y$. The three curves shown
are  the the first, second and third-order spin-wave predictions,
corresponding to  short dashed, long dashed
and solid lines respectively.  \label{fig1} }

\figure{The perturbation diagrams that contribute to the ground state
energy $E_0/N$. The crosses represent the interaction vertices as
indicated; the lines represent boson excitations in the intermediate
states. To save space, we have not differentiated between $\alpha$
and $\beta$ bosons and possible time ordering of the vertices in
the diagrams. \label{fig2} }

\figure{The estimates of $S^2 \Delta E_f^{(-2)}/N$ as a function
of the lattice size $L$ and anisotropy parameter $x$. \label{fig3} }


\newpage
\begin{references}
\bibitem[*]{byline}e-mail address: zwh@newt.phys.unsw.edu.au
\bibitem[$\dagger$]{byline}e-mail address: 87b0090@cumulus.csd.unsw.oz.au
\bibitem{bar}T. Barnes,  Int. J. Mod. Phys. C {\bf 2}, 659(1991).
\bibitem{man}E. Manousakis, Rev. Mod. Phys. {\bf 63}, 1(1991).
\bibitem{neu}H. Neuberger and T. Ziman, Phys. Rev. B {\bf 39}, 2608(1989).
\bibitem{fis}D.S. Fisher, Phys. Rev. B {\bf 39}, 11783(1989).
\bibitem{and} P.W. Anderson, Phys. Rev. {\bf 86}, 694(1952).
\bibitem{kub}R. Kubo, Phys. Rev. {\bf 87}, 568(1952).
\bibitem{ogu}T. Oguchi, Phys. Rev. {\bf 117}, 117(1960).
\bibitem{har}A.B. Harris, D. Kumar, B.I. Halperin, and P. Hohenberg,
Phys. Rev. B {\bf 3}, 961(1971).
\bibitem{kop}P. Kopietz, Phys. Rev. B {\bf 41}, 9228(1990).
\bibitem{cas}G.E. Castilla and S. Chakravarty, Phys. Rev. B {\bf 43},
13687(1991).
\bibitem{iga}J. Igarashi and A. Watabe, Phys. Rev. B {\bf 43},
13456(1991); {\bf 44}, 5057(1991)
\bibitem{can}C.M. Canali, S.M. Girvin, and M. Wallin, Phys. Rev.
 B {\bf 45}, 10131(1992).
\bibitem{goc}I.G. Gochev, submitted to Phys. Rev. B.
\bibitem{cz}C.J. Hamer, W.H. Zheng and P. Arndt,
Phys. Rev. B {\bf 46}, 6276(1992).
\bibitem{atk}K.E. Atkinson, in `An Introduction to Numerical Analysis',
(John Wiley \& Sons, Inc., 1978).
\bibitem{woy}F. Woynarovich and H.P. Eckle, J. Phys. A: Math. Gen.
{\bf 20}, L97(1987); M. Takahashi, Prog. Theo.
Phys. {\bf 50}, 1519(1973); {\bf 51}, 1348(1974).
\bibitem{bet}H.A. Bethe, Z. Phys. {\bf 71}, 205(1931).
\bibitem{hal}F.D.M. Haldane, Phys. Lett. A {\bf 93}, 464(1983);
Phys. Rev. Lett. {\bf 50}, 1153(1983).
\bibitem{aff}I. Affleck, J. Phys.: Condens. Matter {\bf 1}, 3047(1989).
\bibitem{fipr}M.E. Fisher and V. Privman, Phys. Rev. B {\bf 32}, 447 (1985).
\bibitem{gut}A.J. Guttmann, in `Phase Transitions and Critical Phenomena',
 Vol.13, ed. C. Domb and J. Lebowitz (New York, Academic, 1989).
\bibitem{run}K.J. Runge, Phys. Rev. B {\bf 45}, 7229(1992).
\bibitem{sin}R.R.P. Singh, Phys.\ Rev.\ B {\bf 39}, 9760(1989);
{\bf 41}, 4873(1990); R.R.P. Singh and D.A. Huse, Phys.\
Rev.\ B {\bf 40}, 7247(1989).
\bibitem{car}J. Carlson, Phys. Rev. B {\bf 40}, 846(1989).
\bibitem{gro}M. Gross, E. Sanchez-Velasco, and E. Siggia, Phys. Rev. B
 {\bf 40}, 11328(1989).
\end{references}
\end{document}